\documentclass[%
 reprint,
superscriptaddress,
nofootinbib,
 amsmath,amssymb,
 aps,
]{revtex4-2}
\usepackage[utf8]{inputenc}
\usepackage{url}
\usepackage{amsmath}
\usepackage{mathtools}
\usepackage{graphicx}
\usepackage{xcolor}
\usepackage{nicematrix}
\usepackage{comment}
\usepackage{appendix}
\usepackage{hyperref}
\usepackage[capitalize]{cleveref}
\usepackage{tcolorbox}
\usepackage{qcircuit}

\hypersetup{
    colorlinks=true,
    linkcolor=blue,
    filecolor=magenta,      
    urlcolor=cyan,
    pdfpagemode=FullScreen,
    }

\definecolor{LighterGray}{rgb}{235,235,235}
\DeclareMathOperator*{\argmax}{\arg\!\max}
\newcommand{\ket}[1]{|#1\rangle}
\newcommand{\bra}[1]{\langle#1|}
\DeclareMathOperator{\Tr}{Tr}
\begin{document}
\title{Discovery of Optimal Quantum Error Correcting Codes via Reinforcement Learning}
\author{Vincent Paul Su}
\affiliation{Center for Theoretical Physics \& Department of Physics, University of California, Berkeley, CA 94720, USA}

\author{ChunJun Cao}
\affiliation{Joint Center for Quantum Information and Computer Science, University of Maryland, College Park, MD, 20742, USA}
\affiliation{Institute for Quantum Information and Matter, California Institute of Technology, Pasadena, CA, 91125, USA}
\affiliation{Department of Physics, Virginia Tech, Blacksburg, VA, 24060, USA}

\author{Hong-Ye Hu}
\affiliation{Department of Physics, University of California San Diego, La Jolla, CA 92093, USA}
\affiliation{Department of Physics, Harvard University, 17 Oxford Street, Cambridge, MA 02138, USA}
\affiliation{Harvard Quantum Initiative, Harvard University, 17 Oxford Street, Cambridge, MA 02138, USA}

\author{Yariv Yanay}
\affiliation{Laboratory for Physical Sciences, 8050 Greenmead Dr., College Park, MD 20740, USA}

\author{Charles Tahan}
\affiliation{Laboratory for Physical Sciences, 8050 Greenmead Dr., College Park, MD 20740, USA}

\author{Brian Swingle}
\affiliation{Department of Physics, Brandeis University, Waltham, Massachusetts 02453, USA}
\date{May 11, 2023}

\begin{abstract}
The recently introduced Quantum Lego framework provides a powerful method for generating complex quantum error correcting codes (QECCs) out of simple ones. We gamify this process and unlock a new avenue for code design and discovery using reinforcement learning (RL). One benefit of RL is that we can specify \textit{arbitrary} properties of the code to be optimized. We train on two such properties, maximizing the code distance, and minimizing the probability of logical error under biased Pauli noise. For the first, we show that the trained agent identifies ways to increase code distance beyond naive concatenation, saturating the linear programming bound for CSS codes on 13 qubits. With a learning objective to minimize the logical error probability under biased Pauli noise, we find the best known CSS code at this task for $\lesssim 20$ qubits. Compared to other (locally deformed) CSS codes, including Surface, XZZX, and 2D Color codes, our $[[17,1,3]]$ code construction actually has \textit{lower} adversarial distance, yet better protects the logical information, highlighting the importance of QECC desiderata. Lastly, we comment on how this RL framework can be used in conjunction with physical quantum devices to tailor a code without explicit characterization of the noise model.
\end{abstract}
\maketitle

\section{Introduction}

Quantum error correction is a critical step in our quest for fully fault-tolerant quantum computation. Beyond its practical significance in quantum computing, its theoretical aspect also has significant impact in many areas of physics, including condensed matter theory~\cite{Kitaev03} and quantum gravity~\cite{Almheiri:2014lwa,Pastawski:2015qua}. With the recent advances in quantum hardware, the practical and theoretical impacts of quantum error correcting codes (QECC) are rapidly converging on various experimental platforms~\cite{egan2020fault, Cai2021, Egan2021, Krinner2022, Xue2022,RydbergQEC}. 
Given this broad and rapidly developing set of applications, no one-size-fits-all approach can possibly be sufficient. Instead, it is highly desirable to devise and implement flexible and customizable methods for code design which can be tailored to the special needs of each particular application
\cite{RLsurface,XZZX,Dua2022,Roffe2022,VQAQEC}. Here we propose and demonstrate a new approach to custom code generation based on reinforcement learning and the `Quantum Lego' paradigm.

\begin{figure}[htbp]
    \centering
    \includegraphics[width = 0.99\linewidth]{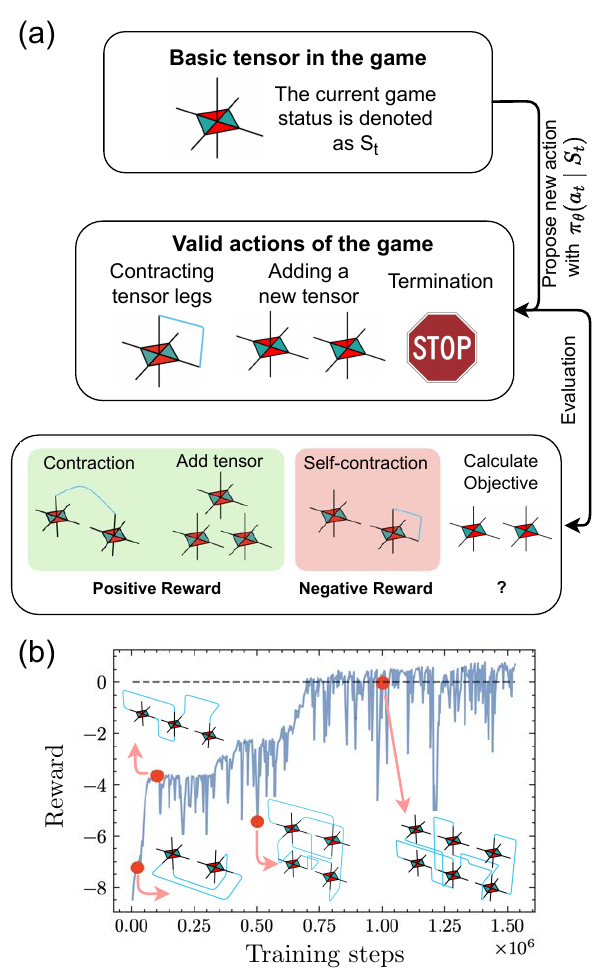}
    \caption{(a) QL game overview. The game starts off with a single copy of the T6 lego. The agent can choose from three actions, adding another T6 lego, choosing legs to contract, or terminating the game. As part of reward shaping, we give a small positive reward for adding another lego, performing a contraction or ending the game. When the contraction connects legs that were previously connected in the tensor network, the agent receives a small negative reward. Code properties are calculated once the game has ended. (b) We show the cumulative rewards the agent earns as it plays the game. A positive final reward (Eq.~\ref{eq:final_reward}) corresponds to outperforming the Color code at protecting against biased noise. Arrows denote repesentative codes generated at different time steps. 
    }
    \label{fig:legogame}
\end{figure}

The Quantum Lego (QL) framework, introduced in Ref.~\onlinecite{Cao:2021ibt} and relatedly \cite{Farrelly22}, provides a systematic method of constructing large quantum error correcting codes in a modular fashion by gluing smaller codes, generalizing code concatenation. As the name implies, these building blocks, or `legos', can be used to create a wide variety of interesting codes. Furthermore, Ref.~\onlinecite{Cao:2021ibt} shows that the framework is universal such that any quantum code can in principle be obtained from just a few types of legos. The landscape reachable by just one type of lego is less clear, but it has been demonstrated through examples that one can construct a variety of non-trivial codes even with this restriction. For example, the rich behavior exhibited by the toric code and the holographic (HaPPY) code can emerge from many copies of the same underlying lego. 

However, beyond code concatenation, there is no known strategy for building up a larger code from the smaller blocks that guarantees an improvement of certain code properties. Previously, this was done either manually, guided by intuition, or through random or exhaustive search~\cite{Cao:2021ibt, Cao22}, using tensor weight enumerator polynomials. In view of the need for codes tailored to particular applications, such manual or exhaustive search strategies do not provide a scalable approach to code design. 

Here, we show that it is possible to gamify the code-building process based on Quantum Legos and teach a machine to construct useful quantum codes. We use a reinforcement learning approach in which an agent manipulates the lego blocks according to simple rules with the goal of maximizing an objective function. The tunability of the objective function makes this approach powerful and versatile. For example, by training on the objective of protecting against skewed noise we were able to find a code that is more robust than any other known code we have compared to. Even stronger results may be obtained in the future from a hybrid approach using interaction with a quantum device to define and calculate the objective function.

We consider reinforcement learning (RL) because it has already achieved beyond human level performance in a variety of areas, including Chess, Go, Poker, and Atari games~\cite{doi:10.1126/science.aar6404,Silver2017,Mnih:2015jgp,doi:10.1126/science.aay2400}. In order to learn these strategies, all that is required is lots of experience, that is, the ability to play the game many times over and over. Within physics, RL has been applied to various problems including the knot problem, the string landscape, quantum circuit compilation, and quantum optimal control~\cite{Gukov:2020qaj,Carifio:2017bov,fosel2021quantum,PhysRevX.12.011059}. RL approaches have also played an important role in optimizing code parameters, for example, improving surface code performance by modifying a base code using lattice surgery~\cite{RLsurface} and exploring the learning of error models and decoders ~\cite{RLdecoder}. 

Our work distinguishes itself from these previous manuscripts in that the goal of learning is to establish a concrete framework that gamifies the QL code building process. By applying RL techniques within this framework, we hope to a) identify moves that generally improve the code properties, b) search for hardware-specific codes given the noise model or access to the quantum device, and c) find new codes with a flexible graph architecture as opposed to a fixed one. We demonstrate the capabilities of this framework by applying it to two tasks. The first task involves finding QL codes that maximize the code distance, of which na\"ive concatenation is an example. Our agent finds a code that achieves the same distance as concatenation with significantly fewer qubits required. Furthermore, it is actually an optimal CSS code with respect to linear programming (LP) bounds~\cite{Grassl:pc,Grasslthesis}. The second task is hardware motivated, where the goal is to minimize the probability of a logical error occurring under a biased noise model. Here, we outperform the best known CSS codes for up to 20 qubits.

The outline of the paper is as follows. In Sec.~\ref{sec:methods}, we provide background on both Quantum Legos and reinforcement learning. Sec.~\ref{ssec:methods_lego} focuses on the modular construction of lego codes and their graphical interpretation as a tensor network. In Sec.~\ref{ssec:methods_rl}, we present how to frame QL code building as a game to be played by an RL agent. In Sec.~\ref{sec:results}, we apply the QL code building framework to two different tasks. We showcase the agent's ability to produce optimal distance CSS codes in Sec.~\ref{ssec:dmax}. In Sec.~\ref{ssec:biased}, we task the agent with building robust codes under a particular, though not fine-tuned, biased noise model. We achieve state of the art results for CSS codes of comparable size. Finally, we conclude with some remarks and future extensions in Sec.~\ref{sec:discussion}.

\section{Methods}\label{sec:methods}

In this section, we describe how to turn Quantum Legos into a game that is amenable to reinforcement learning methods. First, we briefly review how the QL framework can be employed to build new codes in a modular fashion from existing codes. A salient feature is that the design of these codes has a graphical description based on tensor networks. For a more in-depth review of the formalism, see App.~\ref{app:lego_bg} and~\cite{Cao:2021ibt}. 
Then, we turn the problem of learning new codes into a game with a discrete set of actions and states which can be understood by reinforcement learning algorithms. While the rules of the game are fixed, the final reward can be chosen based on desired properties of the resulting code. We will exploit this freedom in Sec.~\ref{sec:results} to learn codes tailored to different objectives.

\subsection{Quantum Legos}\label{ssec:methods_lego}
\begin{figure}
  \centering
  \includegraphics[width=0.9\linewidth]{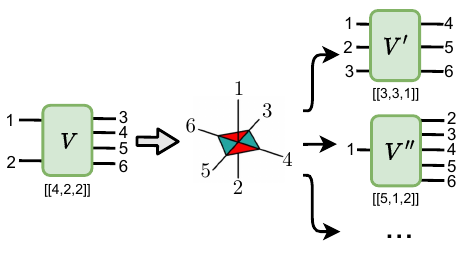}
  \caption{Channel state duality. The encoding map $V$ of a QECC, taking logical qubits to their physical counterparts, can be interpreted as a state on all qubits. The tensor describing the state $\ket{\psi_{V}}$ is simply given by the matrix elements of $V$. When $V$ is the encoding map for the $[[4,2,2]]$ code with stabilizer generators $\langle XXXX, ZZZZ\rangle$, we call this tensor the T6 lego. Note the ambiguity in going from a tensor to a QECC that comes from assigning each of the legs to be a logical or physical qubit. Because of this ambiguity, for example, we could interpret $\ket{\psi_{V}}$ as either a $[[4,2,2]]$ code or a $[[5, 1, 2]]$ code.
  }
 \label{fig:csd}
\end{figure}

Quantum Legos is a modular framework that consists of three primary elements --- the lego set $\mathcal{L}$, connection of the legos in the set, and assignment of logical legs. 

Any lego $\ell\in\mathcal{L}$ is a tensor that can be obtained from the encoding map of a (generally small) QECC. These legos also have a number of symmetries which we can track in addition to their tensor components. For example, one such lego is shown in \cref{fig:csd}, which graphically represents the tensor $\mathcal{V}_{i_1i_2i_3i_4i_5i_6}$ such that each leg is mapped to a tensor index. The tensor components are obtained from the encoding isometry $V:\mathbb{C}^{4}\rightarrow \mathbb{C}^{16}$ of a $[[4,2,2]]$ self-dual CSS code. Its stabilizer group is generated by $\langle XXXX, ZZZZ\rangle$. The logical $X$ and $Z$ operators are given by weight-2 $X$ and $Z$ operators respectively.
In general, when $\ell$ is derived from a stabilizer code, its tensor components are fully fixed by the symmetries it inherits from the parent stabilizer code. Therefore it suffices to track only the symmetries of the lego blocks, instead of its tensor components.\footnote{The lego in this example have symmetries represented graphically in Fig. 17 of~\cite{Cao:2021ibt}.}

To build up larger codes, we can connect the legos by gluing their legs (Fig.~\ref{fig:d4_codes}). This graphical operation corresponds to a tensor contraction where one sums over the indices that correspond to the legs to be glued.\footnote{Physically, the tensor contraction can also be mapped to a Bell fusion~\cite{fbqc}.}
The combined tensor network inherits all symmetries of the individual legos that satisfy the additional constraints imposed by the fused legs. The transformation of the symmetries under such gluing operations is captured graphically as operator pushing or operator matching (App~\ref{app:lego_bg}). Ref.~\onlinecite{Cao:2021ibt} showed that these symmetries fully define the newly formed tensor network, which can then be converted into a quantum code.\footnote{For actual implementation, these graphical moves are represented as operations over the check matrices of the individual lego blocks which only involve simple row and column operations and are thus efficiently implementable. The resulting symmetries of the tensor network over $n$ dangling legs can be encoded in a larger check matrix of size $n\times 2n$.}

Finally, to convert the tensor network into a code, we choose  some legs as logical qubits (inputs) and the remaining as physical (outputs). Such a tensor network then acts naturally as an encoding map. Note that this conversion process is not unique as the tensor (network) corresponds to a class of codes related to each other by the Choi-Jamiolkowski isomorphism. For example, although the tensor in Fig.~\ref{fig:csd} represents a $[[4,2,2]]$ code by construction, it can also be used to define a $[[6,0,3]]$ or a $[[5,1,2]]$ CSS code by choosing the logical legs differently. In later sections, we fix the choice of logical leg to resolve this ambiguity. 

The formalism also furnishes any QL code with a distance verification protocol and an optimal decoder~\cite{Farrelly22,CLG} under i.i.d. single qubit error channels~\cite{CLG}. We do not consider active error correction in this work, but we make use of the quantum weight enumerators made available by~\cite{shorlaflamme, hu2020weight,Cao22,CLG} to compute code distance and to estimate logical error probabilities. When $\mathcal{L}$ consists of only legos from stabilizer codes, the tensor contractions and symmetries are also efficiently trackable via operations on their respective binary matrix representations as shown in Appendix D of Ref.~\cite{Cao:2021ibt}. This determines the stabilizer generators (or check matrix) of the resulting code, from which its unitary encoding circuit can be obtained~\cite{Aaronson}. We review the basics of QL and illustrate several simple code constructions in  Appendix~\ref{app:lego_bg}. For more details, see the original work~\cite{Cao22}.

In the following, we focus on the class of codes that can be constructed from a single type of lego block derived from the self-dual CSS $[[4,2,2]]$ code. We call it a T6 lego and any code constructed using only such blocks a T6 QL code.
Despite this restriction, T6 QL codes contain many types of interesting codes including the toric code, surface code, and trivalent 2D Color codes.

\subsection{Gamifying Quantum Legos\label{ssec:methods_rl}}
Reinforcement Learning (RL) is a subfield of machine learning that learns from experience, e.g.~playing a million games of chess, rather than labeled data sets, e.g.~image classification. With the goal of using machine learning to help with code design, we set out to reframe lego building as a game to be played. To build a game, one needs to specify a set of states $S$ (such as the set of board configurations in chess) and the set of actions $A$ (corresponding to moves to be made). The process of building a QL code can be made atomic with the following sets $S$, $A$.

\begin{tcolorbox}  
\begin{center}
\textbf{Quantum Lego Game}
\end{center}
Given a set of legos $\mathcal{L}$, a state $s \in S$ consists of two pieces of information
\begin{enumerate}
    \item A list of $n$ legos in use $(\ell_1, \ldots, \ell_n), \ell_{i} \in \mathcal{L}$.
    \item A list of $m$ edges $((l_{a_1}, l_{b_1}), \ldots,(l_{a_m}, l_{b_m}))$. Each edge contains a pair of legs $(l_{a_j}, l_{b_j})$ to be contracted.
\end{enumerate}

From a state $s$, there are four possible actions $a \in A$.
\begin{enumerate}
    \item Add a new lego $\ell_{n+1}$.
    \item Contract two dangling legs $(l_{a_{m+1}}, l_{b_{m+1}})$.
    \item Designate a leg as logical.
    \item Terminate.
\end{enumerate}
\end{tcolorbox}

Every time a lego $\ell$ is added, its legs are sequentially numbered. For example, a T6 lego has 6 legs. The first T6 lego would have legs numbered 1 through 6. Once a second lego is added, any two indices between 1 and 12 would specify a valid contraction. The index pair $(a_j,b_j)$ labels the two tensor legs that are connected to form an edge $j$ in the tensor network.

These simple rules are enough to generate all quantum codes given a universal lego set. In principle, this game can be played \textit{ad infinitum}. In practice, we place a few constraints to make the state space finite for computational purposes. For example, we limit the total number of lego blocks and the number of moves the agent can make. The former also constrains the maximum size of the code that can be produced. In this work, we will also fix the choice of logical leg to be the first leg of the first lego in order to streamline the learning process. This leaves us with three actions, adding a new lego to our tensor network, contracting legs in the tensor network, or ending the game. See Fig.~\ref{fig:legogame} for a graphical summary.

When the game is terminated, the final reward is calculated based on the resulting tensor network. In general, the reward could be tied to any code property, such as the distance, the encoding rate, the check weights, etc. 
We emphasize that the training objective for the agent is arbitrary. This flexibility is a useful feature: Any desirable code property that can be quantified can be optimized by the learning algorithm.
In this work, we will consider adversarial distance and the probability of a logical error under a biased noise model. For stabilizer codes, this amounts to looking at the ($X$ and $Z$) weights of logical operators as in Eqs.~(\ref{eq:unnorm_err_prob}) and~(\ref{eq:norm_err_prob}).

We note that ideally the reward should be efficiently calculable, as this calculation will need to be repeated for every iteration of the game, which the agent needs to be able to play many times.
When it comes to quantum codes, this can be a significant limitation, as properties of a quantum system are often exponentially difficult to calculate.
This hurdle can be surmounted by the use of hybrid algorithms~\cite{VQAQEC}, where a classical learner has accesses to a quantum machine for the purpose of calculating the objective function.
In particular, our game could be paired with a quantum computer that performs the actual encoding and decoding functions of the QECC and measures their robustness to induced noise -- or simply to the real noise seen by the machine.

Another common pitfall in RL is the problem of sparse rewards. Much like in a game of chess, a reward that only comes at the end of the long game can make it difficult for the agent to associate good intermediate actions with the final positive rewards. To counteract this,  one can provide small rewards to guide the agent, an idea known as \emph{reward shaping}. In our game, we assign small positive rewards to adding a new tensor ($r_{\text{add tensor}}= .1$), to contracting tensors which were not previously connected in the network ($r_{\text{contraction}}=.05$), and to ending the game ($r_{\text{terminate}} = .15$). Conversely, we assign a small penalty when the agent contracts legs of tensors which are already connected ($r_{\text{self-contraction}} = -.1$) (see \cref{fig:legogame}). Since the distance of a code takes integer values, these rewards were chosen so that taking multiple actions would still be a secondary effect.

When trying to maximize the code distance, these heuristics are important since the majority of moves (self-contractions) when starting with a single T6 lego lead to decreasing code distance. Without reward shaping, the agent will simply maximize its expected reward by avoiding these bad moves by choosing to terminate the game early. Thus, reward shaping can be viewed as coercing the agent to continue playing the distance maximization game. When designing codes for the biased noise model, we actually found the same reward shaping structure to work well.

Contrasting with chess, we point out two aspects of our setup that differ. The first is that ``winning'' is less well-defined. The agent simply tries to get the best score it can, but empirical results can't tell us about whether the agent attains the best possible outcome. The second aspect is that unlike chess where an agent learns to play against an opponent who may take a variety of actions, the code game is deterministic. Thus, the agent only needs to demonstrate a single instance of a valid code to be useful in practical applications.

\section{Results}\label{sec:results}
Here, we present the results of our framework applied to two different objectives for quantum codes:
\begin{enumerate}
    \item Maximizing the code distance,
    \item Minimizing the probability of a logical error under an i.i.d. qubit biased noise model.
\end{enumerate}

To carry out the numerics, we utilize standard RL libraries available in Python. In particular, we use the Gym~\cite{openai_gym} framework for setting up the lego game and Stable-Baselines3~\cite{stable-baselines3} for their implementation of the Masked Proximal Policy Optimization (PPO) learning algorithm~\cite{PPO, MPPO}. For the curious reader interested in RL, see App.~\ref{app:rl_bg} for a brief overview aimed at physicists.

In both tasks, our learning algorithm produces state of the art codes. In the case of distance maximization, our code saturates LP bound for CSS codes and requires a significantly fewer number of qubits to achieve the same distance produced by na\"ive concatenation. For the protection of logical information under biased noise, we find a novel code that outperforms popular CSS codes of comparable size.

\subsection{Maximizing code distance}\label{ssec:dmax}
\begin{figure}
  \centering
  \includegraphics[width=.97\linewidth]{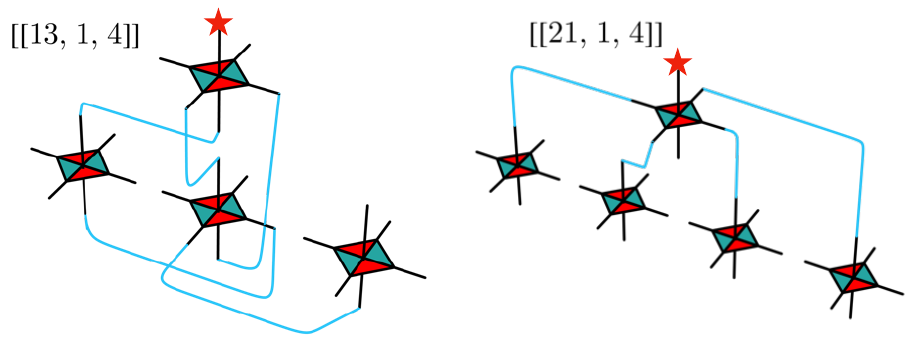}

  \caption{An example of a distance $d=4$ code found in the QL game where the task was to maximize the distance of a single logical qubit. The base distance of the individual T6 legos was two. On the left, we show a $[[13, 1, 4]]$ code that achieves the same distance as the naive concatenation strategy with significantly fewer qubits. For both codes, the red star labels the logical leg.}
 \label{fig:d4_codes}
\end{figure}

As a first demonstration of the power of learning to play with legos, we start by training an agent to maximize the code distance given access to copies of the T6 lego. To fix the ambiguity of going from a tensor network to a code, we will always take the first qubit of the first lego to be the only logical leg.
As a baseline, we compare to na\"ive code concatenation, requiring five copies of the base code, that yields a distance $[[21,1,4]]$ code (Fig.~\ref{fig:d4_codes}).
This code can clearly be constructed as a T6 QL code. Within our QL game approach, we find a $[[13, 1, 4]]$ code that requires significantly fewer physical qubits to implement.

We also note that this $[[13, 1, 4]]$ code saturates the upper bound on distance for CSS codes of this size $n$ and number of logical qubits $k$, making it an optimal code.\footnote{For $k=1$, the smallest CSS code with $d=4$ requires $n=12$ qubits.} In the current era of NISQ devices, finding codes which use fewer physical qubits can be greatly beneficial. Note that the encoding rate was not explicitly part of the game's reward, though one is free to include this as part of the objective.

\subsection{Protecting against biased noise\label{ssec:biased}}

To highlight the flexibility of RL methods for code design, we now turn our attention to a more physically relevant noise model. Given that not all noise is created equal, designing codes that are tailored to particular shortcomings of a device may accelerate reaching fault tolerant computation. A simplified setup where this occurs is if Pauli $Z$ errors are more likely than $X$ errors.

\begin{figure*}
  \centering
  \includegraphics[width=0.93\linewidth]{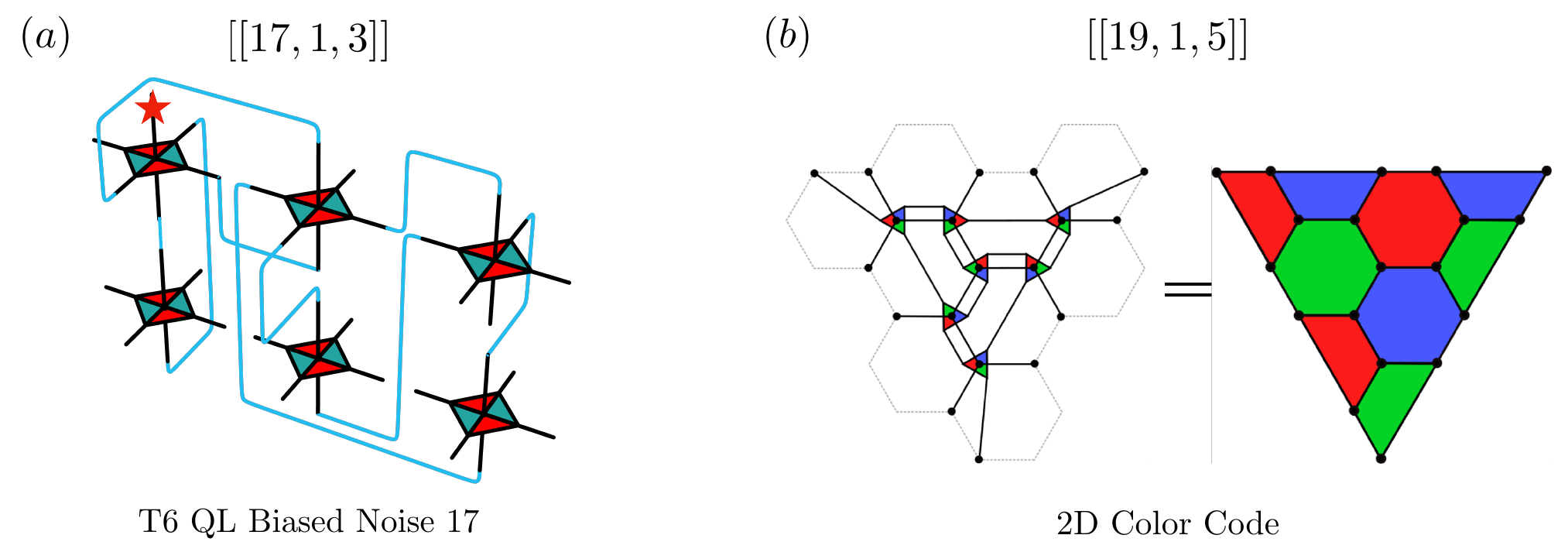}
  \caption{ (a) Lego code trained to minimize logical probability error assuming biased noise $p_{x} = 0.01$, $p_{z} = 0.05$. The red star labels the logical leg. The resulting code parameters are $[[17, 1, 3]]$. Compared with other codes (Table~\ref{tab:error_rates}), this lego code actually has a shorter \textit{adversarial} distance, yet it has a lower probability for a logical error to occur. This probability is calculated by looking at the $X$ and $Z$ weights of logical operators as in Eqs.~(\ref{eq:unnorm_err_prob}) and~(\ref{eq:norm_err_prob}). (b) The $[[19,1,5]]$ trivalent 2D Color code (right). It can be built from encoding tensors of the Steane code (left), each of which is the contraction of two T6 legos. Physical legs or qubits are marked with a solid dot whereas logical legs are suppressed.}
 \label{fig:logical_lego}
\end{figure*}

For such a biased noise model, we use the $[[19,1,5]]$ trivalent 2D Color code (Fig.~\ref{fig:logical_lego}a) as a benchmark since it can also be constructed as an T6 QL code~\cite{CLG}, though its construction requires lego moves that we don't consider in this work.
Strikingly, we find a novel $[[17, 1, 3/4]]$ QL code with $X$ and $Z$ distances $d_X=3, d_Z=4$ that outperforms the 2D Color code, an optimal $[[17, 1, 5/5]]$ CSS code~\cite{CSSlike,Grassl:codetables}, (and other surface code variants) at minimizing the probability of a logical error. We say a code is optimal if its parameters saturate the linear programming bound. Surprisingly, the adversarial distance for our T6 QL code is actually \textit{lower}, meaning that the minimal weight of logical operators is less, than in the comparison codes. Nevertheless, the QL code has many fewer low-weight logical operators (Table~\ref{tab:code_weights}) which results in improved robustness. Note that the lower error rate of the T6 QL code is not a fine-tuned artifact but is robust against perturbation of physical noise rates (Appendix~\ref{app:RLcodes}).  

\begin{table}[]
    \centering

    \begin{NiceTabular}{|c|c|c|c|}[colortbl-like]
    \hline
        Code & $[[n,k,d_X/d_Z]]$ & $\begin{array}{c}p_L\\ (10^{-5})\end{array}$ & 
        $\begin{array}{c}p^{\text{norm}}_L\\ (10^{-5})\end{array}$ \\
        \hline
        \rowcolor{lightgray}T6 BN 13A & $[[13,1,3/4]]$ & $.973$ & $2.16 $ \\
        \rowcolor{lightgray}T6 BN 13B & $[[13,1,3/4]]$ & $.973$ & $2.16 $ \\
        CSS Self-Dual$^{\dagger}$ & $[[13,1,3/3]]$ & $26.8$ & $59.5 $ \\
        \rowcolor{lightgray} T6 DM 13$^{\dagger}$ & $[[13,1,4/4]]$ & $5.68$ & $12.6 $ \\
        \hline
        Reed-Muller & $[[15,1,3/7]]$ & $1.43 $ & $3.61 $ \\
        Surface (4x4) & $[[16,1,4/4]]$ & $1.46 $ & $3.89 $\\
        XZZX (4x4) & $[[16,1,4/4]]$ & $1.07 $ & $2.86 $ \\
        \hline
        \rowcolor{lightgray}
         T6 BN 17 & $[[17,1,3/4]]$ & $\textbf{.404}$ & $\textbf{1.15} $ \\
        CSS self-dual$^{\dagger}$ & [[17, 1, 5/5]] & .726 & 2.06 \\
        2D Color & $[[19,1,5/5]]$ & $.456 $ & $1.46 $ \\
        \hline
        XZZX (4x5) & $[[20, 1, 4/4]]$ & $.665 $ & $2.26 $ \\
        XZZX (5x4) & $[[20, 1, 4/4]]$ & $.665 $ & $2.26 $ \\
        Surface (4x5) & $[[20, 1, 4/5]]$ & $.438$ & $1.48 $ \\
        Surface (5x4) & $[[20, 1, 5/4]]$ & $6.58$ & $22.2 $ \\
        \hline
        \end{NiceTabular}
    \caption{Error rates for best known CSS codes when $p_x=0.01,p_z=0.05$. BN and DM refer to the tasks that the agent was trained on, biased noise and distance maximization. Under our biased noise model, we compute the normalized an unnormalized logical error rates for a variety of codes. For the learning procedure, we asked the agent to minimize the normalized error rate and it outperformed all other CSS codes of similar size. Codes marked with $^{\dag}$ denote codes that are optimal with respect to linear programming bounds on the distance.
    }
    \label{tab:error_rates}
\end{table}
\begin{table}[]
    \centering

    \begin{NiceTabular}{|c|c|c|c|c|}[colortbl-like]
    \hline
        Code & $[[n,k,d]]$ & wt $3$ & wt $4$ & wt $5$ \\
        \hline
         \rowcolor{lightgray} T6 BN 17 & $[[17, 1, 3]]$ & 4 & 12 & 32\\
        2D Color code & $[[19, 1, 5]]$ & 0 & 0 & 108 \\
        \hline
        \end{NiceTabular}
    \caption{Distribution of logical operator sizes. Under i.i.d. qubit noise, low weight logical operators are more likely than high weight operators to corrupt the logical information. Comparing our lego code to the 2D Color code, we find that our code has a lower minimal weight (e.g. smaller $d$), yet achieves an overall lower probability of logical error.}
    \label{tab:code_weights}
\end{table}

The above observations suggest that code design for asymmetric noise is a difficult problem to analyze by hand, yet for the algorithm it amounts to a simple tweak of the reward function. Instead of the reward being tied to the adversarial distance, we give the agent a positive reward for a lower logical error probability. Let us describe precisely the quantity we compute.

In this example, we consider biased noise models with independent bit flip and phase error probabilities $p_x$ and $p_z$, respectively. Note that these T6 QL codes admit a number of decoders~\cite{CLG}, such as the maximal likelihood decoder~\cite{Farrelly22}, for which the corresponding error probabilities can be computed or estimated. Although such decoders are easy to incorporate into the framework, they are more time-consuming to implement. Instead, we use two simplifying estimates for logical error probability.  The first one is given by the probability of an undetectable logical error $p_L$, which is the probability that a non-trivial logical operation has been applied to the encoded state due to bit flip and phase flip errors on individual qubits.  In other words, this is the lower bound for the logical error rate regardless of the choice of decoder. We refer to it as the \emph{unnormalized error rate}, which is related to error detection. A measure of similar properties is used in~\cite{VQAQEC}. Because bit flip and phase errors occur independently, for each non-trivial logical Pauli operator $\bar{L}$ such that the $X$-weight $wt_X(\bar{L})=w_x$ and $Z$-weight $wt_Z(\bar{L})=w_z$, the total probability that it comes from a physical error is $p(w_x,w_z)=p_x^{w_x}p_z^{w_z}(1-p_x)^{n-w_x}(1-p_z)^{n-w_z}$. Therefore,
\begin{equation}\label{eq:unnorm_err_prob}
p_L = \smashoperator{\sum_{w_x,w_z=0}^n} C_{w_x,w_z} p(w_x,w_z),
\end{equation}

where $C_{w_x,w_z}$ enumerates the number of non-identity logical operators with the same $X$ and $Z$ weights.

For the second measure, we consider a detection-based decoder which resets the system upon measuring a non-trivial syndrome. This is the logical error probability \textit{given} that one measures trivial error syndrome, meaning no decoding needs to be applied. The corresponding logical error probability is
\begin{equation}\label{eq:norm_err_prob}
    p_L^{\text{norm}}=p_L/p_{s=0},
\end{equation}
where $p_{s=0}$ is the probability of measuring trivial syndromes (assuming no measurement errors).
We refer to the logical error rate obtained this way as the \emph{normalized error rate}.
In training for this task, we keep the same game as in the previous distance maximization game, but simply ask the agent to minimize this normalized error rate.

Both measures can be computed exactly using the double quantum weight enumerator polynomials~\cite{hu2020weight}.
Further details of these enumerators and how they are used to compute the error probabilities are explained in App.~\ref{app:qwep}.

For benchmarking purposes, we compare with the $[[19,1,5/5]]$ 2D Color code and other existing codes of similar sizes, such as the XZZX code, surface code, and Reed-Muller code (Table~\ref{tab:error_rates}). Since the 2D Color code provides the lowest probability for a logical error and can be obtained as a T6 QL code, we tie the final reward of the agent to its performance relative to the 2D Color code
\begin{equation}\label{eq:final_reward}
    r_f = -\log\left(\frac{p^{\text{norm}}_{L, {\rm lego}} }{p^{\text{norm}}_{L, {\rm 2D Color}}} \right).
\end{equation}
Hence, the agent receives a positive reward when its error probability is lower than that of the 2D Color code.

Remarkably, we find a $[[17,1,3/4]]$ code that outperforms these codes at both the normalized and unnormalized error probabilities. We note that the stabilizer weight distribution of the basic T6 lego building block is symmetric with respect to $X$ and $Z$. However, the RL agent has identified contractions that break this symmetry, producing a code that is tailored to this biased noise problem. Notably, the learning procedure also yielded two T6 QL BN codes at $n=13$ with comparable performance. They are asymmetric $[[13,1,3/4]]$ CSS codes which is near optimal~\cite{CSSlike}. These codes outperform their counterparts at $n=13$ despite having a lower distance than some. The reason is again related to the operator weight distribution. See Fig.~\ref{fig:13qubitXZ} and Appendix~\ref{app:qwep} for details. Interestingly, the two $n=13$ codes have distinct scalar enumerators, indicating that they are not simply related by local unitary operations or simple permutation of the physical qubits. However, they have the same double weight enumerators, hence yielding the same logical error probabilities under biased noise.

In Fig.~\ref{fig:legogame}, we plot the cumulative reward of the learning algorithm over time for on this biased noise task. Intermediate rewards are given by the reward shaping structure mentioned in Sec.~\ref{ssec:methods_rl} and the final reward is given by Eq.~\ref{eq:final_reward}. Without reward shaping, a positive reward corresponds to beating the color code (and other surface code variants). For the agent shown in Fig.~\ref{fig:legogame}, the maximum number of legos was 7 and the number of actions was limited to 20.

\section{Discussion}\label{sec:discussion}

Our work highlights a promising avenue for the discovery of new quantum codes. The use of model-free RL algorithms underscores the power to design codes with targeted purposes. Here we considered a simple biased error model, but one could easily tailor the reward function to take into account hardware-specific information. 
Note that our approach is not limited to the scale of qubit codes we have here. With quantum legos, we merely keep track of the symmetries of the tensor network, as opposed to working in the Hilbert space of all the qubits. Thus, as long as there is an efficient contraction for the tensor network representing the QL code, all steps in our framework can be carried out efficiently.

\subsection{Towards hardware tailored codes}
Ultimately, the best kind of QECC is one that is tailored~\cite{Dua2022,VQAQEC} to the device it is run on. A number of approaches for doing this have been considered and implemented for small scale devices of a few qubits. Here we provide another approach based on the Quantum Lego game. The approach is more scalable compared to the existing methods thanks to its modular architecture and use of generalized concatenation, and can accommodate a greater variety of codes as the game allows for a high degree of flexibility in the network architecture. We give a road map towards integration with near-or-intermediate term quantum hardware in the preliminary stage.

The reward function can often be difficult to evaluate when it is tied to certain code properties. Depending on the type of code, this cost can scale rapidly with system size. For sufficiently large systems (which may yet be beyond our current hardware capabilities) it may be more advantageous to use the quantum device to evaluate the cost function directly. Instead of benchmarking our devices, determining the best-fit error models, then finding a good code based on that model, our strategy in this case will be to obtain a code tailored to the hardware while being oblivious to the error model.

Note that for each code that the agent finds (or starts with), we are given its check matrix and the corresponding tensor network. Using this information, one can prepare a particular encoded state on the quantum device. For a circuit based preparation scheme, one can easily obtain a 2-local Clifford encoding circuit from the check matrix with $O(n^2/\log(n))$ complexity \cite{Cleve1997}. For a platform that is measurement or fusion based, we prepare the individual lego blocks unitarily then perform Bell fusion to create the requisite tensor network, followed by the standard post-measurement correction protocols. This process is constant depth plus classical processing. 

With an encoder circuit available, any number of objective functions can be evaluated. An estimate of memory error can be obtained by simply waiting some fixed amount of time, or single- and multi-qubit gate errors can be similarly estimated by acting with the equivalent logical operators on the encoded system. This can be followed by syndrome measurements,  decoding and final read out of the logical information. By repeating this process a large number of times, we can estimate the logical error probability, which will determine the reward we feedback to the agent.

A number of decoders can be used. 
The most trivial, but not very practical, decoder would measure the syndromes, discard states with non-trivial values, and measure the logical state of those with trivial syndromes to see if they match the target state. This provides an estimate for the normalized error probability discussed in Sec.~\ref{sec:results}. Such a method is more feasible for a code when the noise levels are low such that trivial syndromes occur with exceedingly high probability. A more sophisticated decoder can also be implemented using a model provided by~\cite{Farrelly22,CLG}. For instance, with an initial guess of the error model, one can also use the maximum likelihood tensor network decoder which works for all Quantum Lego codes. However, their output will depend on benchmarks or guesses about the error model, which then needs to updated to achieve true optimality. Perhaps a ML approach to finding a decoding schedule in conjunction with the encoding scheme is ultimately needed for best outcomes (e.g.~\cite{RLdecoder,NNdecoder}), but we leave that for future work.

\subsection{Other Modifications}
The framework can be easily modified to include other user requests and to speed up the learning process/reduce learning cost. It can also be used in conjunction with other existing methods. We comment on a few possibilities in this section. 

\textit{Decent Quantum LDPC codes:} 
While we have focused on $k=1$ codes by maximizing distance or minimizing logical error probability, encoding rate and stabilizer check weights are also key deciding factors for the practicality of a quantum code. Therefore, an obvious extension of the current model is to include rate or check weights also into the cost function. 

The former can be extended by choosing more than one logical leg. However, the task of finding the right tensor legs as logical inputs can also be difficult by brute force. The current game is played with a singular designated logical input. This is a simpler, but more restricted, version of QL as it allows one to identify multiple legs as logical inputs. To unlock the full potential of the framework that allows the construction of known examples such as the surface code or 2d Color code,  one can incorporate the move of identifying a tensor leg as logical input. Together with the moves we have in Fig~\ref{fig:legogame}, they completely characterize the actions allowed by QL and is sufficient for recreating all quantum codes.

Check weights can be trickier as the naive RREF of a check matrix does not necessarily yield the optimal weight especially with spatially local constraints. However, it can be a useful starting point until a better method is found. Note that~\cite{Cao:2021ibt} identified moves in a case by case basis that can keep the check weights low. A successful RL model in this extension has the promise help us identify more general moves for reducing stabilizer check weights while also increasing the code distance. Progress in this direction would be a key result in identifying good moves beyond code concatenation.

\textit{Transfer learning:}
Access to a quantum device can be expensive and it can take a large number of iterations before the agent converges on a good strategy. While the above section describes a framework which works in principle, it is far more realistic in practice to first train the agent classically with respect to a known error model that approximates the behaviour of the real quantum device. After a certain amount of training for the agent to acquire a basic understanding of the game, we then continue the training where the cost function is now provided by a real quantum device in a process like the one described above. This allows the agent to make relatively minor policy updates which can reduce the cost and runtime on the quantum computer while achieving a similar outcome on error suppression.

\textit{Approximate cost functions:}
A major resource consuming task in our current model lies in the choice of our cost functions. For both of the examples, we computed distance or weight enumerators exactly. Because such a problem is NP-hard, the general cost can be as expensive as $O(\exp(d))$ even with the tensor network methods in~\cite{Cao22,CLG}. However, such level of accuracy may not be needed for practical code designs. To that end, it is reasonable to substitute the exact values of logical error probabilities with approximate values using Monte Carlo methods or approximate tensor network contractions. For instance, MPO-based approximate contraction schemes achieves $O(n\log n)$ compared to $O(\exp(\sqrt{n}))$ for a planar tensor network~\cite{2021Chubb} like the one for the surface code~\cite{2014MPO}. For intermediate term systems with larger sizes, this can be a practical approach.

\textit{Fixed architecture codes:}
In this work we have reduced the complexity by connecting only one type of lego. However, this is ultimately a limiting factor, as other blocks are needed to obtain other more flexible codes. However, this can drastically expand the already vast state space and thus the complexity for learning and search. A method to reduce the state space is to consider a more rigid architecture instead of any graph with bounded degree like the one we have here. For instance, we can insist that the agent must connect blocks in manners that produces a planar graph or the native connectivity required by the quantum hardware. On the most restrictive case, we can fix the graph architecture completely and only allowing the agent to choose different legos for each vertex. This is somewhat reminiscent of the VQA-based strategy~\cite{VQAQEC} but with discrete range of choices.

\textit{Compatibility with VQA based methods:}
 For smaller sized devices, it is possible to prepare individual legos with VQA~\cite{VQAQEC} that are close to optimal. These codes are generally not Pauli stabilizer codes. However, the same VQA-based approach is not directly scalable. One way to circumvent this bottleneck is to use code concatenation. We can further extend the scalability of this framework by treating each small-sized VQA code as a lego. As the QL framework remains valid for all quantum codes, one can then combine the VQA legos using our RL setup. Furthermore, the tensor network is furnished with an efficient method for computing the quantum weight enumerators~\cite{Cao22,CLG}, which help characterize the properties of non-stabilizer quantum codes. This is a unique advantage of the QL based method, which unlike other current frameworks, can easily accommodate non-stabilizer codes.
\section*{Note Added}
A few days after our work was released on the arXiv, a related work, Ref.~\cite{FarrellyRL}, that uses reinforcement learning to create tensor network codes also appeared. There they focus on optimizing adversarial distance using non-CSS legos.   

\section*{Acknowledgement}
We thank Haowei Deng for pointing out a typo, Markus Grassl for the helpful comments on linear programming bounds for CSS codes and Robert Huang for his helpful suggestions. C.C. acknowledges support from the U.S. Department of Defense and NIST through the Hartree Postdoctoral Fellowship at QuICS and the National Science Foundation (PHY-1733907). C.C. and B.G.S. acknowledge support from the Air Force Office of Scientific Research (FA9550-19-1-0360). The Institute for Quantum Information and Matter is an NSF Physics Frontiers Center. V.P.S. gratefully acknowledges support by the NSF Graduate Research Fellowship Program under Grant No. DGE 1752814, the DOE Office of Science under QuantISED Award DE-SC0019380. H.Y.H. is grateful for the support from the Harvard Quantum Initiative Fellowship.
\bibliography{main}

\pagebreak

\appendix

\section{Review of Quantum Lego Formalism}\label{app:lego_bg}
In this section, we review more of the formalism behind quantum legos. For more details, please see~\cite{Cao:2021ibt}. As in the main text, our analysis will mostly follow a single building block, called the T6 lego, that describes a $[[4,2,2]]$ self-dual CSS code whose stabilizer group is $\langle XXXX, ZZZZ\rangle$. The logical $X$ and $Z$ operators are given by weight-2 $X$ and $Z$ operators respectively. With the T6 lego, we illustrate the main conceptual tools of Quantum Legos.

\subsubsection{Channel-State Duality}
The rules for combining these modular codes are inspired by tensor networks and will have a graphical construction. To make this connection, we make use of the \textit{channel state duality}, wherein we can reinterpret the encoding map $V$, from the two logical qubits to the four physical qubits, 
\begin{equation}
    \mathcal V = \smashoperator{\sum_{i_1,\dotsc,i_6}}V_{i_1i_2i_3i_4i_5i_6}\ket{i_3,i_4,i_5,i_6}\bra{i_1,i_2} 
\end{equation}
as a quantum state,
\begin{equation}
    \ket{\psi_{V}} = \smashoperator{\sum_{i_1,\dotsc,i_6}}V_{i_1i_2i_3i_4i_5i_6}\ket{i_1,i_2,i_3,i_4,i_5,i_6}.
\end{equation}
Matrix elements of $V$ are simply the wave function coefficients for basis elements on both input and output legs, which are now on equal footing. See Fig.~\ref{fig:csd}. Note that going from a state to a channel is not necessarily unique, as there is a freedom in choosing which (and how many) of the legs to choose as logical. For example, we could interpret the tensor arising from the $[[4, 2, 2]]$ code as a $[[5, 1, 2]]$ code,
\begin{equation}
    \mathcal V^{\prime} = \smashoperator{\sum_{i_1,\dotsc,i_6}}V_{i_1i_2i_3i_4i_5i_6}\ket{i_2,i_3,i_4,i_5,i_6}\bra{i_1}.
\end{equation} To standardize the notation, let us denote the input legs as $1$ and $2$, with the output legs corresponding to $3,4,5,6$.

For stabilizer codes, the resulting state $\ket{\psi_{V}}$ is a stabilizer state. The stabilizers of that state e.g.~${S : S\ket{\psi_{V}}= +1 \ket{\psi_{V}}}$ are formed by taking $O^{\rm (L)}\otimes O^{\rm (P)}$ where $O^{\rm (L)}$ denotes a logical Pauli operator and $O^{\rm (P)}$ denotes the corresponding physical operator. To go from the state back to a code, one can isolate a set of physical legs $P'$, and refactor $S = O^{\rm (L)}\otimes O^{\rm (P)} = O^{\rm (L')}\otimes O^{\rm (P')}$. {If $O$ is not a Pauli operator, one should take extra care in this conversion by taking the complex conjugate such that a stabilizer for the dual quantum state is given by ${O^{\rm (L)}}^*\otimes O^{\rm (P)}$, but for Pauli matrices $O=O^{*}$ up to global phase factors.}

\subsubsection{Operator Pushing}
Ultimately, we would like to construct larger codes via tensor networks of tensors constructed in the previous section. The notion of \textit{operator pushing} is a useful way to keep track of how errors propagate in a code. For a tensor network code, one can find all stabilizer generators by performing operator pushing on a network. Here, we illustrate operator pushing on a single tensor. 

\begin{figure}
  \centering
  \includegraphics[width=0.9\linewidth]{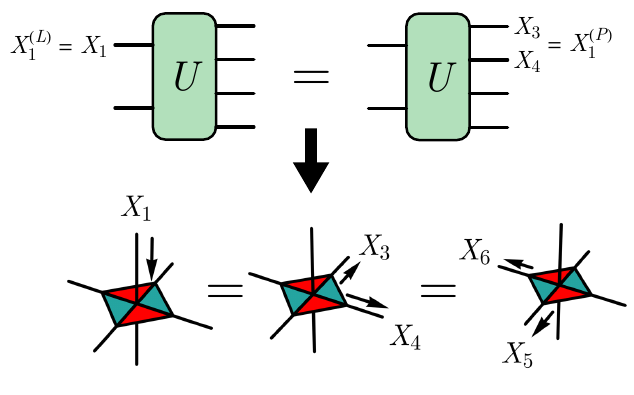}
  \caption{Operator pushing. An operator $O^{\rm (L)}$ acting on a logical leg has a corresponding implementation, $O^{\rm (P)}$ on the physical legs. In terms of the tensor representation, one can visualize this as pushing $X_{1}$ ``through'' the tensor to an equivalent operator $X_{3}X_{4}$. The action of both simultaneously $O^{\rm (L)}\otimes O^{\rm (P)}$ is a stabilizer of the resulting state. For Stabilizer codes, there are many equivalent representations of $O^{\rm (P)}$ related by the action of the stabilizer group. In the diagram $X_{3}X_{4}\sim X_{5}X_{6}$ where $\sim$ denotes logical equivalence due to the stabilizer $X_{3}X_{4}X_{5}X_{6}$.
  }
  \label{fig:op_push}
\end{figure}

Consider the application of a Pauli $X$ operator on the first logical qubit, $X^{\rm (L)}_{1} = X_{1}I_{2}$. In this code, it is physically implemented by applying $X^{\rm (P)}_{1} = X_{3}X_{4}I_{5}I_{6}$,
\begin{equation}
    \mathcal V X^{\rm (L)}_{1} \ket{i_1i_2}=
    X^{\rm (P)}_{1}\mathcal V \ket{i_1i_2} . 
\end{equation}
In the tensor network language, this can be thought of as pushing an operator through the tensor, $X^{\rm (P)}_{1}\mathcal V = \mathcal V X^{\rm (L)}_{1}$, see Fig.~\ref{fig:op_push}. Using the channel-state duality, we immediately find
\begin{equation}
    (I^{\rm (L)}\otimes X^{\rm (P)}_{1})\ket{\psi_V} = 
    (X^{\rm (L)}_{1}\otimes I^{\rm (P)})\ket{\psi_V},
\end{equation}
making it clear that $\ket{\psi_{V}}$ lies in the +1 eigenspace of $X^{\rm (L)}_{1}\otimes X^{\rm (P)}_{1} = X_1I_2X_3X_4I_5I_6$, making it one of the code stabilizers.

To summarize, given $O^{\rm (L)}$ and $O^{\rm (P)}$ of a code with encoding $V$, $O^{\rm (L)} \otimes O^{\rm (P)}$ yields a stabilizer of the resulting state $\ket{\psi_{V}}$. This is a direct consequence of the channel-state duality. Graphically, this allows us to perform the pushing, where any operator can be acted on by $O^{\rm (L)} \otimes O^{\rm (P)}$ to yield an equivalent operator on other legs.

\subsubsection{Gluing Legos}

\begin{figure}
  \centering
  \includegraphics[width=\linewidth]{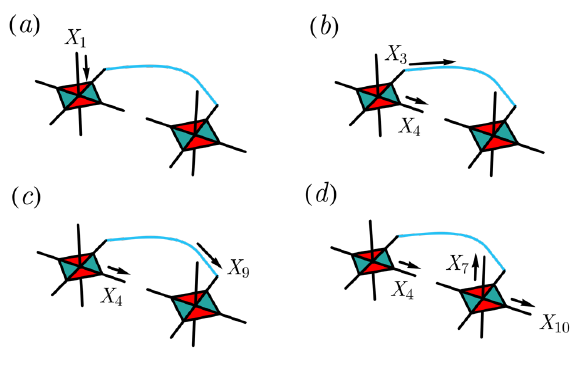}
  \caption{Single-trace operation via operator flow. Given two copies of T6, we glue the legs of two tensors by projecting onto a maximally entangled state (blue). The stabilizer of the resulting state can be found by performing operator pushing. a) We start with $X_{1}^{\rm (L)} = X_{1}$ acting on the first qubit. b) Pushing through the first tensor yields $X_{3}X_{4}$. c) Since qubits 3 and 9 are entangled, $X_{3}\sim X_{9}$. d) Finally, pushing through the second tensor, we get $X_{1}^{\rm (P)}=X_{4}X_{7}X_{10}$, so $X_{1}^{\rm (L)}\otimes X_{1}^{\rm(L)} = X_{1}X_{4}X_{7}X_{10}$ is a stabilizer of the resulting tensor network after contraction.
  }
  \label{fig:single-trace}
\end{figure}

Let us take two copies of this code, labeling the legs of $A$ as 1-6 and $B$ as 7-12 with the convention for numbering the same as before. Consider contracting legs 3 and 9, which amounts to a Bell fusion, or a projection on the maximally entangled state. We would like to understand the stabilizers for the remaining ten qubit state. Suppose we take stabilizers of the individual tensors $X_{1}^{\rm (L)} \otimes X_{1}^{\rm (P)} = X_{1}I_{2}X_{3}X_{4}I_{5}I_{6}$ and $X_{7}^{\rm (L)} \otimes X_{7}^{\rm (P)} = X_{7}I_{8}X_{9}X_{10}I_{11}I_{12}$, then $X_{1}I_{2}X_{4}I_{5}I_{6}X_{7}I_{8}X_{10}I_{11}I_{12}$, would be a stabilizer of the remaining qubits. Here we make use of the fact that $X^{2}= I$. An alternate way to understand this construction of the resulting stabilizer is to push $X_{3}$ ``through'' the other tensor. This is illustrated in Fig.~\ref{fig:single-trace}.

In constructing the full set of stabilizers, we simply perform this operation for all operators acting on the glued legs, taking care to ensure operators acting on the traced legs are matched. Although naively the total number of stabilizer matchings is exponential in the system size, in the case of stabilizer codes, there is an efficient algorithm to keep track of the glued codes via a check matrix operation known as conjoining (Appendix D of \cite{Cao:2021ibt}). 

Note that any two legs may be glued together. When two legs of the same tensor (more generally, a connected component) are glued, we refer to it as a self-trace. Otherwise, we refer to the operation as a single-trace.

\begin{figure}
  \centering
  \includegraphics[width=0.9\linewidth]{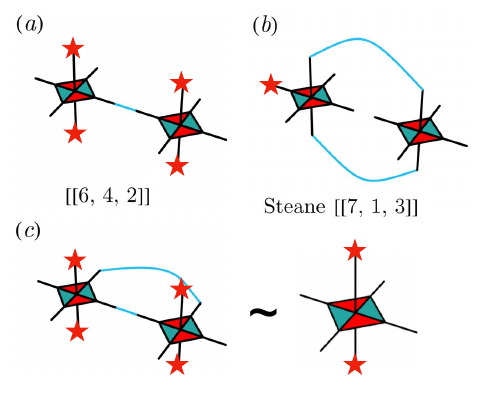}
  \caption{Examples of code construction using Quantum Lego. Given just two copies of the T6 lego, we already see a variety of behavior. (a) A single contraction yields a code that encodes more physical qubits with a better encoding rate, at the same distance as the original code. (b) Picking a single logical leg after tracing two legs reproduces the well-studied [[7, 1, 3]] Steane code. (c) Performing two traces on different choices of legs reduces to a single copy of the [[4,2,2]]. This demonstrates the rich and unexpected behavior that these simple contractions can yield.
  }
  \label{fig:example_codes}
\end{figure}

In general,~\cite{Cao:2021ibt} showed that tracing together stabilizer codes yield another stabilizer code. If the individual legos are (self-dual) CSS codes, then up to re-definitions via the Choi-Jamiolkowski isomorphism, so are the resulting codes obtained from gluing their physical legs. Note that although $[[4,2,2]]$ is self-dual, the $[[5,1,2]], [[6,0,3]]$ codes from the same tensor are not, because of weight-3 stabilizers. Furthermore, tracing of two non-self-dual codes can be self-dual, e.g. the Steane code example in Fig.~\ref{fig:example_codes}b. Therefore, the codes we obtain in this work are CSS codes, but the self-duality property need not be preserved during the game.

\subsubsection{Examples of Code Constructions}

\begin{figure}
    \centering
    \includegraphics[width=0.6\linewidth]{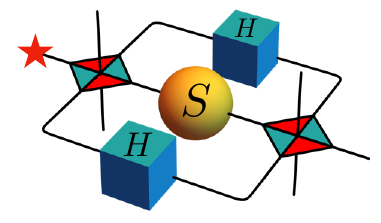}
    \caption{ $[[5,1,3]]$ from two $[[4,2,2]]$ codes with $H$ and $S$ which are the tensors for Hadamard and phase gates. The red star labels the logical leg.
    }
    \label{fig:513}
\end{figure}

With even two copies of the T6 lego, we can see how different lego codes can arise. Performing a single trace between specific legs generates a $[[6, 4, 2]]$ code. Performing two gluing operations can lead to either the $[[7, 1, 3]]$ Steane code, or the original $[[4, 2, 2]]$ code after identifying gauge constraints. The holographic HaPPY code can be constructed from many copies of the perfect $[[5, 1, 3]]$ code. The perfect property of these codes allows for easy operator pushing, which was part of the original inspiration. These examples are obtained by contractions of isometric tensors. Although they are different from naive code concatenation, they can be understood as a concatenated code with suitable applications of the CJ-isomorphism. However, the lego constructions also go beyond concatenation. A simple example is the $[[5, 1, 3]]$ perfect code, which can be constructed with two copies of the T6 lego with 3 contractions if one allows for modifications by $H$ and $S$ gates (Fig.~\ref{fig:513}). 
Other constructions that are built beyond code concatenation include Bacon Shor codes, the 2d Color code, and the Toric code, where one must contract non-isometric tensors. 
Although it might appear restrictive to use only the T6 lego, it is worth mentioning that all examples above except the $[[5,1,3]]$ code can be built from the T6 legos alone. 

\section{Reinforcement Learning Overview}\label{app:rl_bg}
Since reinforcement learning (RL) is not a part of the standard toolkit for physicists, we summarize two learning algorithms that were used through this work, real-time dynamic programming (RTDP) and Masked Proximal Policy Optimization (MPPO). The details of the learning algorithms aren't essential for the paper, but this may provide a useful starting point for those wishing to extend our results. Ultimately, we were able to produce all lego codes in this paper with MPPO, but for pedagogical purposes, we include a discussion on both. For a more in-depth treatment, see these original papers~\cite{rtdp,PPO,MPPO} as well as some classic references on RL~\cite{silver2015, sutton_barto}. 

There are two main quantities that play a distinguished role in learning algorithms, the \textit{value function} $V(s)$, which measures the expected reward of continuing the game from a given state $s$, and the \textit{policy} $\pi(a|s)$, which characterizes the agent's probabilities to take an action $a$ in a given state $s$. Generically, the output of the learning algorithm will be a policy $\pi$ that maximizes the expected reward of playing the game.

\subsubsection{Real-Time Dynamic Programming}
Real-Time Dynamic Programming (RTDP) falls under the general class of algorithms that tries to learn the value function $V(s)$. If one truly has access to $V$, then it is simple to construct the optimal policy $\pi^*(a|s)$ by simply selecting the action $a$ that leads to the highest value successor state $s'$. 
Under mild assumptions, $V(s)$ can be iteratively solved by the Bellman equations, which state that the value of a state is related to the value of states $s'$ which are reachable from $s$. This is akin to numerically solving Laplace's equation by averaging neighboring points.
\begin{equation}
  \label{eq:bellman-iter}
  V^{(\pi)}(s) =  \gamma \sum_{a} \pi(a|s) \left(r_{a} + \sum_{s'}T^{a}_{ss'} V^{(\pi)}(s') \right)
\end{equation}
where $r_{a}$ is the reward for taking action $a$, $\gamma$ is a discount factor, and $T^{a}_{ss'}$ is the probability for the action $a$ to take the agent from state $s$ to $s'$. Note that the value of a state depends on the relative probabilities of visiting neighboring states, hence the dependence on $\pi$.

Due to the recursive structure of the Bellman equations, one can efficiently (polynomial in the number of states) solve this problem via \textit{dynamic programming}. Once $V(s)$ is fully known, it is then easy to back out the policy by simply selecting the action which leads to the successor state $s'$ with the highest value as follows
\begin{equation}
  \label{eq:greedy-policy}
  \pi^{(\text{new})}(a|s) = \delta_{aa'} \, , \, a' \equiv \argmax_{a} r_{a} + \sum_{s'}T^{a}_{ss'} V^{(\pi)}(s')
\end{equation}

In practice, since $V$ depends on the current policy $\pi$, one performs the updates to $V$ and $\pi$ holding the other fixed. In general, there are strong guarantees that this will converge to the optimal policy $\pi^{*}$ after multiple rounds~\cite{sutton_barto}.

While this is conceptually clean and efficient, there is a problem for games in which the number of possible states grows intractably large. The insight behind RTDP is to perform a Bellman type update for only a subset of states which actually appear when playing the game. The agent simply plays the game according to its current policy $\pi$ and keeps track of what states it has seen. Those states are the only ones updated by the Bellman iterations. Note that there is incomplete information about the value function, so we lose the theoretical guarantees. However, in practice, it was sufficient for finding more efficient $d=4$ codes.

\subsubsection{Masked Proximal Policy Optimization}

The other general RL approach is known as \textit{policy learning}, where one attempts to model the function $\pi(a|s;\theta)$ directly with some parameters $\theta$. Commonly, these are parameters for some neural network that model the policy. In contrast to RTDP and other value-based methods, one is agnostic about the inherent value of states and simply tries to tweak these parameters to obtain a large expected reward. There are usually with auxilliary penalties in the objective function to ensure training stability and convergence. The algorithm we employ is the Masked Proximal Policy Optimization (MPPO), a form of policy learning that was recently popularized for its effectiveness and simplicity~\cite{PPO,MPPO}.

Let us define the expected total reward under the policy $\pi_{\theta}$ as $J(\theta)$.
\begin{equation}
  \label{eq:3}
  J(\theta) = \sum_{s}d^{(\pi_{\theta})}(s) V^{(\pi_{\theta})}(s)
\end{equation}
Here, $d^{(\pi_{\theta})}(s)$ is the stationary distribution of a Markov chain with actions selected according to $\pi_\theta$. The stationary distribution refers to the likelihood of being in state $s$ if the agent is allowed to play the game infinitely long under the policy $\pi_\theta$. Thus, $J$ is a weighted sum of the values of those states. This is a formal object that depends on $\theta$ which we will not have access to directly since we are not trying to learn the value function $V$. However, we can get estimates of $J$ by simply playing the game many times.

Since our goal is to maximize the expected returns, captured formally by $J$, one may naturally be inclined to apply gradient ascent with respect to the policy parameters $\theta$. However, there is a thorny conceptual issue. Choosing the correct step size is difficult due to the implicit dependency of both $d$ and $V$ on these parameters. Choose too small a rate and your agent will fail to learn. On the other hand, if the step size is too large, the actions may change drastically, raising convergence issues.

Proximal Policy Optimization (PPO) improves on this naive approach by penalizing large changes in the policy (e.g. if $r \equiv \frac{\pi_{\theta + \Delta\theta}(a|s)}{\pi_{\theta}(a|s)}$ deviates significantly from $1$). Effectively, this amounts to a maximum tolerance for the amount the policy can change. In prior works, one had a penalty for the Kullback-Lieber divergence between $\pi_\theta$ and $\pi_{\theta + \Delta\theta}$. However, the approach of PPO was found to be empirically more effective. $\pi_\theta$ and $\pi_{\theta + \Delta\theta}$. However, the approach of PPO was found to be empirically more effective. For more details see~\cite{PPO}. 

Finally, the last tweak to the learning algorithm is to use \textit{action masking}. Most RL algorithms assume that the set of actions is the same for all times (such as the four cardinal directions if playing Pacman). In more complex games such as ours and chess, the set of valid actions is much larger and not all of them will be applicable in a given state. For our lego game, the set of actions includes the ability to contract any two legs. However, if one of the legs is already contracted, this would be an invalid move. To get around this, one could simply assign a negative reward for making such an improper move and eventually the agent would learn not to try to contract legs that are already contracted. However, it is more sample efficient simply to mask those invalid moves upfront~\cite{MPPO}.

\section{Quantum weight enumerator polynomials}\label{app:qwep}
For each i.i.d. single qubit error, one can determine the undetectable logical error probability from the weight enumerator polynomials of the code \cite{CLG}. To compute the logical error probabilities used in Sec.~\ref{sec:results} under biased noise, let us define the quantum double weight enumerators $A(\mathbf{u}), B(\mathbf{u})$

\begin{align*}
    A(\mathbf{u}) &=\frac{1}{K^2}\sum_{E\in \mathcal{P}^n}\Tr[E\Pi]\Tr[E^{\dagger}\Pi]\mathbf{u}^{f(E)}  \\
    B(\mathbf{u}) &=\frac{1}{K}\sum_{E\in \mathcal{P}^n}\Tr[E\Pi E^{\dagger}\Pi]\mathbf{u}^{f(E)},
\end{align*}
 where $\Pi$ is the projection onto the code subspace, $\mathcal{P}^n$ is the Pauli group over $n$ qubits, and $K$ is the dimension of the code subspace. The variable $\mathbf{u}=(x,y,z,w)$ is a 4-tuple and $\mathbf{u}^{f(E)}= x^{wt_X(E)} y^{n-wt_{X}(E)} z^{wt_Z(E)} w^{n-wt_Z(E)}$. Respectively, $wt_X, wt_Z$ are $X$ and $Z$ weights of the Pauli string $E$. These polynomials are related by a MacWilliams identity
 
\begin{equation}
    A(x,y,z,w)=\frac 1 K B(w+z,w-z,\frac{x+y}{2},\frac{y-x}{2}).
\end{equation}
and contain important information of the code, such as its distance and error detection probability. See~\cite{hu2020weight} for details. Note the normalization convention is different from that of \cite{Cao22}.
 
We can rewrite the enumerators in the following form
\begin{align}
    A(x,y,z,w) &= \sum_{w_x,w_z=0}^n A_{w_x,w_z} x^{w_x}y^{n-w_x}z^{w_z}w^{n-w_z} \\
    B(x,y,z,w) &= \sum_{w_x,w_z=0}^n B_{w_x,w_z} x^{w_x}y^{n-w_x}z^{w_z}w^{n-w_z} 
\end{align}

 Let $\mathcal{E}_{w_x,w_z}$ be the set of Pauli operators that have $X$ weight $w_x$ and $Z$ weight $w_z$. Then for a stabilizer code with stabilizer $\mathcal{S}$, normalizer $\mathcal{N}(\mathcal{S})$,  $A_{w_x,w_z}= |\mathcal{E}_{w_x,w_z}\cap \mathcal{S}|$ and $B_{w_x,w_z}= |\mathcal{E}_{w_x,w_z}\cap \mathcal{N}(\mathcal{S})|$, i.e., up to a known normalization constant, they count the number of stabilizer and normalizer elements with $X$ and $Z$ weights $w_x,w_z$. Therefore, the coefficient $C_{w_x,w_z}$ in (\ref{eq:unnorm_err_prob}) can be written as $C_{w_x,w_z} = B_{w_x,w_z}-A_{w_x,w_z}$. 
 
\begin{figure*}[ht]
    \centering
    \includegraphics[width=\linewidth]{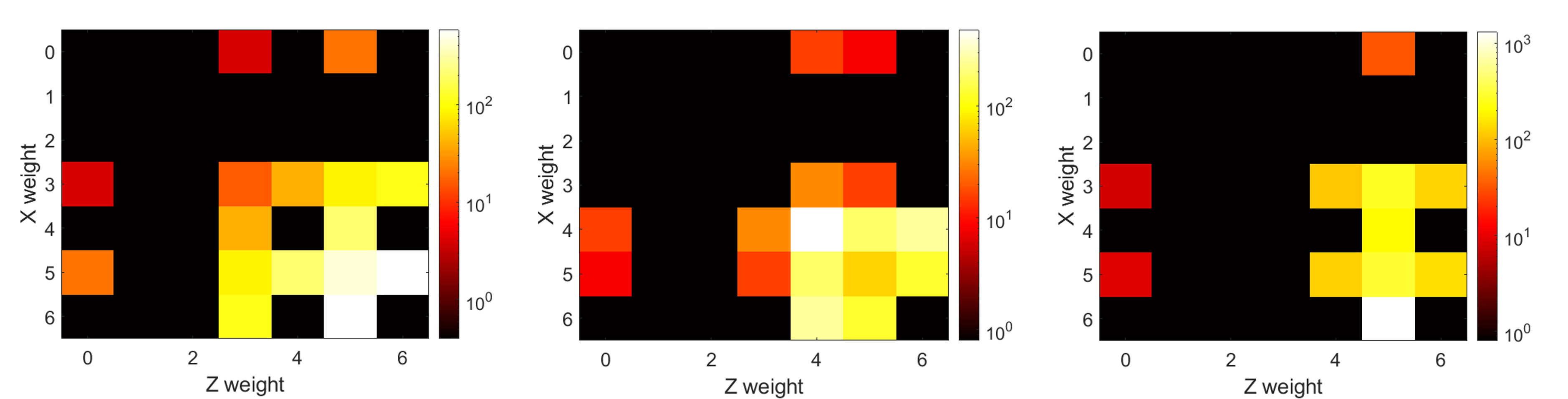}
    \caption{$X$ and $Z$ weight distributions of the non-trivial logical operators for $n=13$ qubit codes. Left: $[[13,1,3]]$ optimal self-dual CSS code. Middle: $[[13,1,4]]$ optimal CSS code. Right: $[[13,1,3/4]]$ T6 QL code.}
    \label{fig:13qubitXZ}
\end{figure*}

 Consider a biased i.i.d. single qubit error where bit flip errors and phase errors occur independently with probabilities $p_X$ and $p_Z$. Then by evaluating the polynomials at $\mathbf{u_p}=(p_x,1-p_x,p_z,1-p_z)$, we see that $$p_{s=0}=B(\mathbf{u}_p)$$ returns the exact probability that the physical errors implements any logical operation (including logical identity) and $A(\mathbf{u}_p)$ computes the probability that the error corresponds to a member of the stabilizer group, which implements a logical identity. Assuming noiseless syndrome measurements, it is clear that $$p_L=B(\mathbf{u}_p)-A(\mathbf{u}_p).$$ They are uncorrectable even in principle because such errors are undetectable. This is the \textit{unnormalized} logical error in the main text, which lower bounds the logical error rate regardless of the choice of decoder. 
Similarly, the \textit{normalized} error probability is given by $$p_L^{norm} = p_L/p_{s=0} = 1-\frac{A(\mathbf{u}_p)}{B(\mathbf{u}_p)}.$$ This is the probability of non-trivial logical error given that one measures trivial syndrome. Physically, this corresponds to a ``decoding''  process where one resets the system whenever an error is detected.

Given a stabilizer group, one can always compute these polynomials by enumerating all relevant stabilizer and normalizer elements. The current-best algorithm for general stabilizer codes scales as $O(2^{n-k})$, where one can enumerate all stabilizer group elements to obtain $A(\mathbf{u})$. One then acquires its dual $B(\mathbf{u})$ using a MacWilliams transform which is at most $O(n^3)$. However, if a Quantum Lego construction is known, then~\cite{Cao22,CLG} allows for a more efficient algorithm that can provide up to exponential speed-ups. As the codes we consider are relatively small in this work, it is sufficient to use the brute force algorithm. For scalability, the tensor network method~\cite{CLG,Cao22} may be used for exact error probability. Alternatively, conventional Monte Carlo methods can also be used where one would compute approximate, instead of exact, error probabilities. 

The check matrices and codes for generating the relevant quantum weight enumerators can be found in App.~\ref{app:RLcodes}. 
Here we also provide the Shor-Laflamme enumerators for the codes listed in Table \ref{tab:code_weights}, such that the coefficients of $C(z)=B(z)-A(z)$ are precisely the weight distribution of the non-trivial logical operators.

\begin{align*}
    &C(z)_{[[17,1,3]]}=4z^3+12z^4+32z^5+144z^6+556z^7\\
    +&1876z^8+5296z^9+12640z^{10}+25020z^{11}+38292z^{12}\\
    +&43712z^{13}+36944z^{14}+22228z^{15}+8396z^{16}+1456z^{17}
\end{align*}

\begin{align*}
    &C(z)_{[[19,1,5]]}=108z^5+765z^7+11406z^9+71523z^{11}\\
    +&252000z^{13}+321363z^{15}+120582z^{17}+8685z^{19}
\end{align*}

For $n=13$, the $[[13,1,3/4]]$ T6 QL code still outperforms other benchmarks at the same size even though it does not beat the color code which requires more physical qubits. 
One may note that its performance is better than the $[[13,1,4]]$ code that is optimal with respect to the LP bound, which has a higher adversarial distance. Even though all three such codes have similar number of low weight non-trivial logical operators, their distributions (Fig.~\ref{fig:13qubitXZ}) are very different, with the QL code having fewer operators that have low $Z$ weights.

\section{Calculating Objective Functions with Quantum Circuits\label{app:Qcircuits}}

As mentioned in the main text, the versatility of the objective function naturally lends itself to implementation with hybrid quantum algorithms. Here, we give a simulated example, calculating undetectable logical error $p_L$ under biased noise, that was used in \cref{ssec:biased}.

A hybrid objective function would consists of several parts, summarized below:
\begin{tcolorbox}  
\begin{center}
\textbf{Hybrid Objective Function Calculation}
\end{center}
Given a state of the game
\begin{enumerate}
    \item Generate an encoder circuit
    \item Run one or more test circuits, of the form:
    \begin{enumerate}
        \item Prepare some single-qubit state
        \item Encode it into the $n$-qubit system
        \item Apply some set of gates to generate noise or errors
        \item Measure stabilizers
        \item Apply a decoder circuit
        \item Measure the final state
    \end{enumerate}
    \item Calculate a cost function based on measurement outputs
    \item Return the result to the classical learner algorithm
\end{enumerate}
\end{tcolorbox}

The encoder circuit mentioned in the first step, which takes in one qubit in an unencoded state and $n-1$ qubits in the $\ket{0}$ state and outputs the encoded $n$-qubit state. This can be done either directly via measurements of the stabilizers, or via a deterministic polynomial algorithm \cite{Cleve1997}. Another approach, which we take here, is to use the building block T6 lego itself and convert the various possible permutations of incoming and outgoing legs into unitary circuits; this can be done whenever there are more outgoing than incoming legs by adding $\ket{0}$-initialized qubits. These sub-circuits can then be connected in the same way that the tensors were contracted by the learner.

This encoder circuit can then be concatenated into a larger circuit in the second step. The initialization of 2(a) can sweep over e.g.~a set of Pauli states or randomly rotated qubit states. Step 2(c) is where the strength of the hybrid approach lies. By applying real operations, from idling the system for a set time to applying logical $X$ or $Z$ gates, one can probe the true noise parameters of the system, allowing the learner to optimize the code for the actual hardware in use. Alternately, one quantum computer can be used to generate codes for another by artificially simulating a known noise model. These sort of calculations, that go beyond stabilizer states and Clifford operations, are exponentially difficult to calculate with classical hardware but remain polynomial with quantum devices.

Here, we give the results of a specific calculation set up to recreate the results of Table \ref{tab:error_rates}. We apply the steps above as follows:
\begin{enumerate}
    \item For the encoder, we use the approach described above, combining several smaller circuits representing different input-output ordering of the H-Lego tensor and connect them as per the contractions of the state.
    \item
    \begin{enumerate}
        \item To capture $X$ and $Z$ errors, we initialize the system in a state $\ket{\pm Y}\otimes \ket{0}^{\otimes n-1}$, where ${\hat Y\ket{\pm Y} = \pm\ket{\pm Y}}$.
        \addtocounter{enumii}{1} 
        \item To recreate the noise model, we apply a single idle gate to each qubit, with chance $p_{x}=0.01$ of an $X$ error and chance $p_z=0.05$ of a $Z$ error.
    \end{enumerate}
    \item We count the number of instances, $N_{s=0}$, where the stabilizers all measured $+1$, including the number of instances where the original state was measured to be different than the initial, $N_{L}$. 
    We return the logical error rate $p_{L}=N_{L}/N_{\rm shots}$ and normalized error rate $p_{L}^{\rm norm}=N_{L}/N_{s=0}$.
\end{enumerate}

\begin{figure}[htb]
    \centering
    \mbox{
\Qcircuit @C=.9em @R=0.1em @!R {
& {\text{(a)}}
& {\text{(b)}}
& {\text{(c)}} \gategroup{2}{4}{5}{4}{.7em}{--}
& {\qquad\text{(d)}} \gategroup{2}{5}{6}{6}{.7em}{--} &
&  {\text{(e)}}
& {\qquad\text{(f)}} \gategroup{2}{8}{5}{9}{.7em}{--}
\\ \lstick{\ket{0}} & \gate{\hat U_{0}} 
    & \multigate{3}{\rotatebox{90}{Encoder}} & \gate{I_{\rm n}} 
    & \multigate{3}{\rotatebox{90}{Stabilizers}} &\qw
	& \multigate{3}{\rotatebox{90}{Decoder}} 
    & \gate{\hat U_{0}^{\dagger}} & \meter 
\\  &  & \ghost{\rotatebox{90}{Encoder}} & \gate{I_{\rm n}} 
    &\ghost{\rotatebox{90}{Stabilizers}} & \qw
	& \ghost{\rotatebox{90}{Decoder}} & \qw & \meter 
\\ & \lstick{\ket{0}^{\otimes n-1}} 
    & \ghost{\rotatebox{90}{Encoder}} & \gate{I_{\rm n}} &\ghost{\rotatebox{90}{Stabilizers}} 
	& \qw & \ghost{\rotatebox{90}{Decoder}} & \qw & \meter 
\\ &  & \ghost{\rotatebox{90}{Encoder}} & \gate{I_{\rm n}} 
    &\ghost{\rotatebox{90}{Stabilizers}} & \qw
	& \ghost{\rotatebox{90}{Decoder}} & \qw & \meter 
\\ & \lstick{\ket{0}^{\otimes n-1}} &  \qw & \qw 
    & \targ \qwx[-1]&  \meter
}}
    \caption{Circuit diagram for the hybrid calculation described in \cref{app:Qcircuits}. $\hat U_0$ rotates the initial $\ket{0}$ state into $\ket{\pm Y}$. Each $I_{\rm n}$ designates a noisy idle gate. Section (c) could be replaced by a more general noisy circuit to generate other cost functions.
    }
    \label{fig:circuitdiag}
\end{figure}
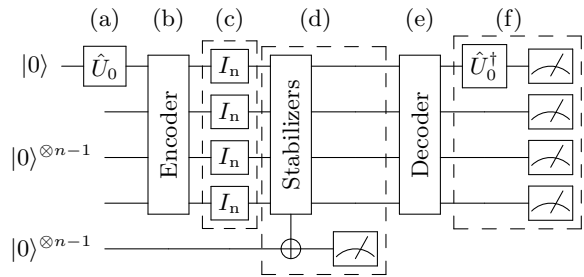

The circuit diagram for this calculation is shown in \cref{fig:circuitdiag}. We perform the simulation using Qiskit \cite{Qiskit}. The results are summarized in \cref{tab:error_rates_qiskit}. We note that we obtain numbers quite similar to those calculated analytically in \cref{tab:error_rates}. Thus the hybrid approach could be readily substituted into the calculation of the cost function. While there is no advantage to doing so with simulated circuits, the use of real devices would allow a dramatic speedup in the calculation time and so the learning time. In general the required time would scale polynomially in the number of qubits, and would be inversely linear to the code error rate.

\begin{table}[]
    \centering

    \begin{NiceTabular}{|c|c|c|c|c|c|}[colortbl-like]
    \hline
        Code & $N_{L}$ & $\begin{array}{c}N_{s=0} \\ (10^{5})\end{array}$ & 
        $\begin{array}{c}N_{\rm shots} \\ (10^{5})\end{array}$ & $\begin{array}{c}p_L\\ (10^{-5})\end{array}$ & 
        $\begin{array}{c}p^{\text{norm}}_L\\ (10^{-5})\end{array}$ \\
        \hline
        \rowcolor{lightgray}T6 BN 13A & $13$ & $4.44$ & $10$ & $1.3$ & $2.9$  \\
        \rowcolor{lightgray} T6 DM 13A & $24$ & $1.78$ & $4$ & $6$ & $13.5$ \\
        \hline
        \end{NiceTabular}
    \caption{Error rates for Lego CSS codes when ${p_x=0.01},{p_z=0.05}$, calculated using Qiskit. Compare with the analytically calculated error rates shown in \cref{tab:error_rates}.
    }
    \label{tab:error_rates_qiskit}
\end{table}

\section{RL Codes}\label{app:RLcodes}
Here we give the check matrices $H=H_X\oplus H_Z$ of the CSS codes made from T6-legos. As usual, each row of the check matrix corresponds to a stabilizer generator. Let $r_P^i\in \mathbb{F}_2$ be the $i$th component of a row vector in $H_P$ where $P=X,Z$. Then its corresponding stabilizer generator is $S=\bigotimes_{i=1}^n P^{r_P^i}$.

For the $[[13,1,4]]$ (DM 13) code by maximizing adversarial distance,
\begin{align*}
H_X^{\text{DM 13}}&=
\left[
\begin{array}{ccccccccccccc}
1&0&0&1&0&1&1&0&0&0&1&1&0\\
0&1&0&1&0&1&0&1&0&0&0&0&0\\
0&0&1&1&0&0&1&1&0&0&1&1&0\\
0&0&0&0&1&1&1&1&0&0&0&0&0\\
0&0&0&0&0&0&0&0&1&0&0&1&1\\
0&0&0&0&0&0&0&0&0&1&1&1&1
\end{array}
\right]\\
    H_Z^{\text{DM 13}}&=
    \left[
\begin{array}{ccccccccccccc}
1&0&0&1&0&0&1&1&0&0&0&1&1\\
0&1&0&1&0&1&1&0&0&0&0&0&0\\
0&0&1&1&0&1&0&1&0&0&0&1&1\\
0&0&0&0&1&1&1&1&0&0&0&0&0\\
0&0&0&0&0&0&0&0&1&0&1&0&1\\
0&0&0&0&0&0&0&0&0&1&1&1&1
\end{array}
\right].
\end{align*}

For the $[[13,1,3]]$ (BN 13A) code found by minimizing normalized logical error probability,
\begin{align*}
    H_X^{\text{BN 13A}}&=
    \left[
    \begin{array}{ccccccccccccc}
1&0&0&0&1&0&1&1&0&0&1&0&1\\
0&1&0&0&1&0&0&1&1&0&1&1&0\\
0&0&1&0&1&0&0&1&1&0&0&1&1\\
0&0&0&1&1&0&1&0&1&0&0&0&0\\
0&0&0&0&0&1&1&1&1&0&0&0&0\\
0&0&0&0&0&0&0&0&0&1&1&1&1
\end{array}
\right]\\
H_Z^{\text{BN 13A}}&=
\left[
\begin{array}{ccccccccccccc}
1&0&0&0&1&0&1&0&1&0&0&0&0\\
0&1&0&0&1&0&1&0&1&0&0&1&1\\
0&0&1&0&1&0&1&0&1&0&1&0&1\\
0&0&0&1&1&0&1&1&0&0&0&0&0\\
0&0&0&0&0&1&1&1&1&0&0&0&0\\
0&0&0&0&0&0&0&0&0&1&1&1&1
\end{array}
\right].
\end{align*}

For BN 13B code,
\begin{align*}
    H_X^{\text{BN 13B}}&=
    \left[
    \begin{array}{ccccccccccccc}
    1&0&0&0&1&0&0&0&0&0&1&0&1\\
0&0&1&0&1&0&1&0&1&0&1&1&0\\
0&1&0&0&1&0&0&1&1&0&0&1&1\\
0&0&0&1&1&0&1&1&0&0&0&1&1\\
0&0&0&0&0&1&1&1&1&0&0&0&0\\
0&0&0&0&0&0&0&0&0&1&1&1&1
\end{array}
\right]\\
H_Z^{\text{BN 13B}}&=
\left[
\begin{array}{ccccccccccccc}
1&0&0&0&1&0&0&0&0&0&1&1&0\\
0&1&0&0&1&0&0&0&0&0&1&0&1\\
0&0&1&0&1&0&1&0&1&0&1&0&1\\
0&0&0&1&1&0&0&1&1&0&1&0&1\\
0&0&0&0&0&1&1&1&1&0&0&0&0\\
0&0&0&0&0&0&0&0&0&1&1&1&1
\end{array}
\right].
\end{align*}

Finally, the BN 17 $[[17,1,3]]$ asymmetric code that minimizes the same logical error,

\begin{align*}
    H_X^{\text{BN 17}}&=
    \left[
    \begin{array}{ccccccccccccccccc}
    1&0&0&0&0&0&0&1&1&0&0&1&1&0&1&0&1\\
0&1&0&0&0&0&0&0&0&0&1&0&1&0&1&0&1\\
0&0&1&0&1&0&0&0&0&0&0&1&1&0&0&0&0\\
0&0&0&1&1&0&0&0&0&0&0&1&1&0&1&1&0\\
0&0&0&0&0&1&0&1&0&0&1&1&0&0&0&1&1\\
0&0&0&0&0&0&1&0&1&0&1&1&0&0&0&1&1\\
0&0&0&0&0&0&0&0&0&1&1&1&1&0&0&0&0\\
0&0&0&0&0&0&0&0&0&0&0&0&0&1&1&1&1
\end{array}
\right]\\
H_Z^{\text{BN 17}}&=
\left[
\begin{array}{ccccccccccccccccc}
1&0&0&0&0&0&1&1&0&0&0&0&0&0&0&0&0\\
0&1&0&0&0&0&0&1&1&0&0&1&1&0&0&0&0\\
0&0&1&0&0&0&0&0&0&0&1&0&1&0&0&1&1\\
0&0&0&1&0&0&0&1&1&0&0&0&0&0&1&1&0\\
0&0&0&0&1&0&0&1&1&0&1&0&1&0&1&0&1\\
0&0&0&0&0&1&1&1&1&0&0&0&0&0&0&0&0\\
0&0&0&0&0&0&0&0&0&1&1&1&1&0&0&0&0\\
0&0&0&0&0&0&0&0&0&0&0&0&0&1&1&1&1
\end{array}
\right].
\end{align*}

\begin{figure*}[ht]
    \centering
    \includegraphics[width=0.95\textwidth]{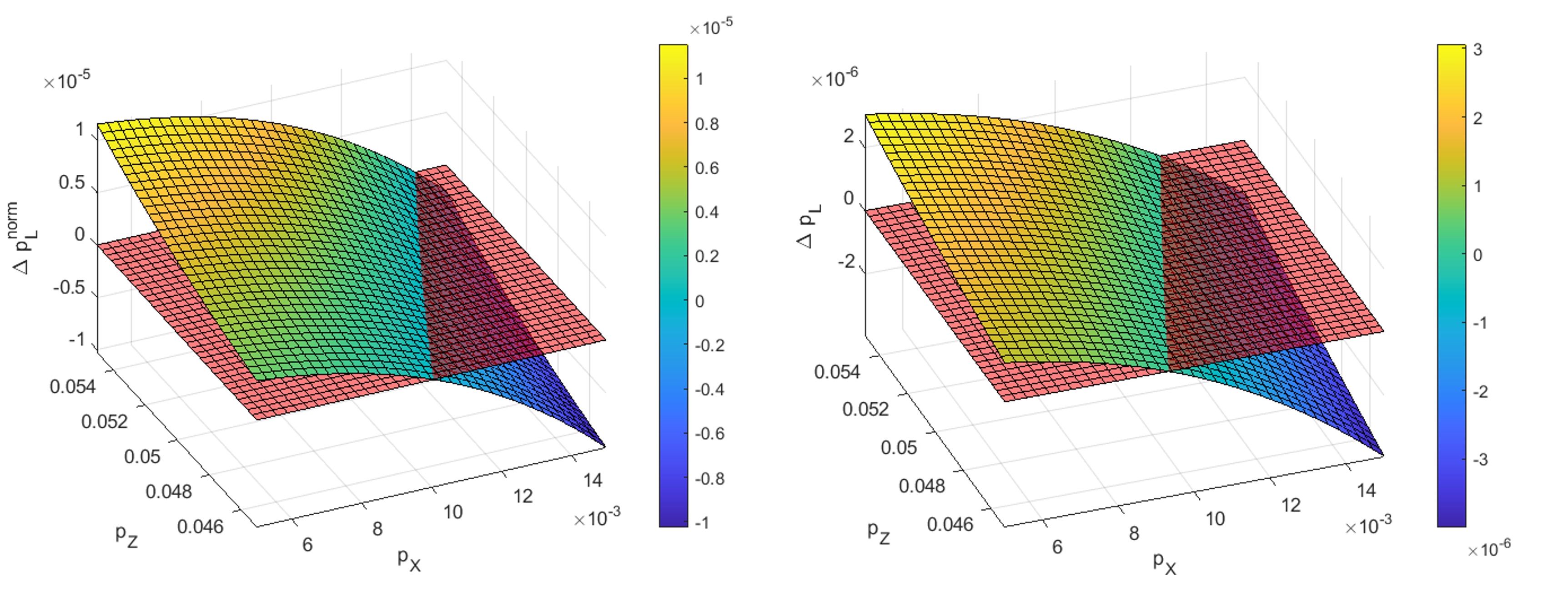}
    \caption{The difference in normalized logical error rate $\Delta p_L^{\rm norm} = p^{\text{norm}}_{L, {\rm 2D Color}} - p^{\text{norm}}_{L, {\rm BN 17}} $ as a function of $p_X,p_Z$ (left) and that of unnormalized error rate (right). The translucent plane in red marks $\Delta p_L=0$ above which the $[[19,1,5]]$ code has higher error rate.}
    \label{fig:perturb}
\end{figure*}

We remark that although the codes found here attains the best known logical error rate for some fixed physical error rate, the code must also not be fine tuned to that particular error probability as error rates from realistic settings are not exact. 
Fig~\ref{fig:perturb} shows that the BN 17 $[[17,1,3]]$ code continues to outperform the $[[19,1,5]]$ Color code in reducing logical error even if one perturbs the physical error probabilities in the neighbourhood of $p_X=0.01, p_Z=0.05$. Therefore the notion of optimality is not fine-tuned.

\end{document}